\documentclass{ws-procs9x6}

\begin{document}

\title{Results from the Commissioning Run of the CMS Silicon Strip Tracker}

\author{Dorian Kcira~\footnote{Supported by the Belgian Interuniversity Attraction Pole, PAI, P6/11.}}

\address{Universit\'e catholique de Louvain, Center for Particle Physics and Phenomenology, Louvain-La-Neuve, Belgium. dorian.kcira@cern.ch}

\begin{abstract}
Results of the CMS Silicon Strip Tracker performance are presented as obtained in the setups where the tracker is being commissioned.
\end{abstract}

\keywords{Silicon Strip Tracker, Compact Muon Solenoid, CERN, HEP}

\bodymatter

\section{Introduction}

The CMS Silicon Strip Tracker~\cite{bib:striptracker}, referred from here on as the Tracker, was commissioned using cosmic muons in two main setups: the Magnet Test Cosmic Challenge (MTCC) and the Tracker Integration Facility (TIF)~\cite{bib:tkintegration}. A large number of studies of the tracker performance and offline data analysis has been performed and many others are still ongoing. Only few of these studies for both setups are presented here.

\section{Signal Properties}

\begin{figure}
\begin{center}
\psfig{file=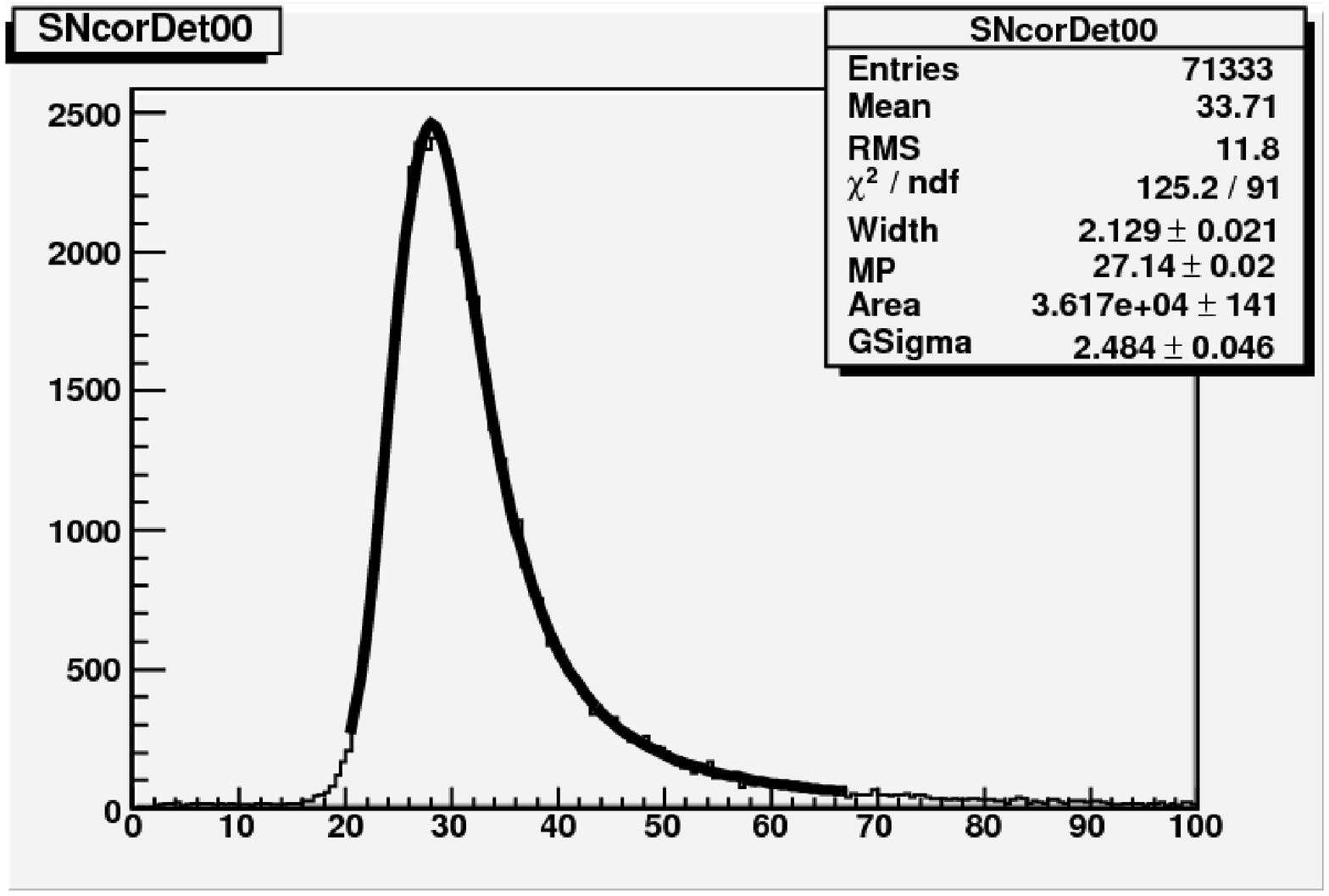,width=2.2in}
\psfig{file=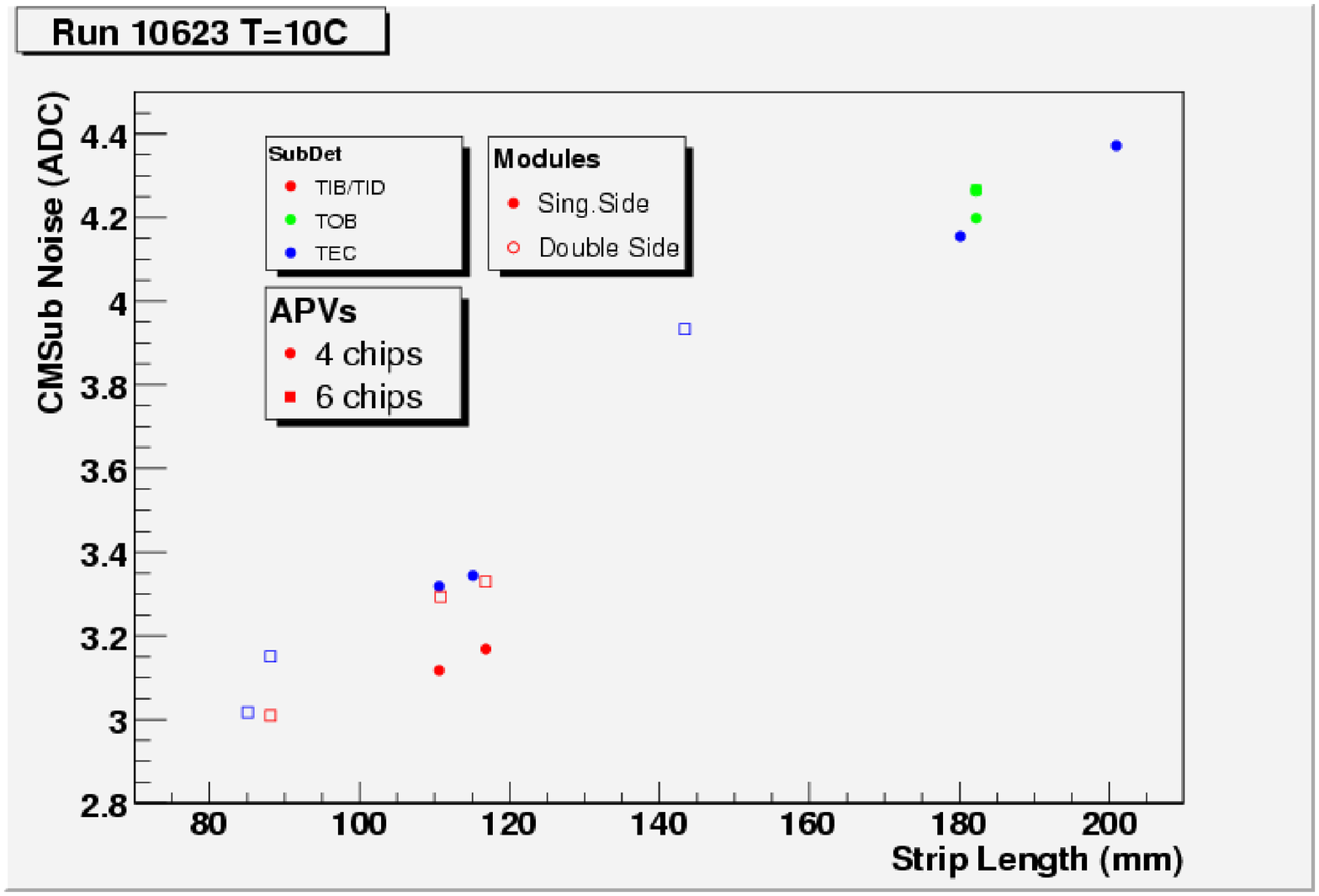,width=2.2in}
\end{center}
\caption{Left plot: Signal to Noise ratio for TIB. Right plot: Noise versus strip length.}
\label{fig:ston}
\end{figure}

In figure~\ref{fig:ston} (right) the noise in ADC counts is plotted versus the length of the micro strips for different subsets of the detector modules at the TIF. As expected, this dependence is linear and the dependence on other  effects, like number of readout chips per module, is small.
The signal-to-noise ratio ($S/N$) for hits associated to reconstructed cosmic muon tracks (section~\ref{sec:tracking}) at the TIF was measured and is shown in figure~\ref{fig:ston} (left) for the TIB. The achieved (most probable) values for the $S/N$ are very good, and are all larger than 27 for all substructures of the tracker. For the TIF data taking the slower readout mode of the chip was used as the trigger timing was not precise enough to use the fast one. Using the fast readout mode will lead to a reduction by a factor of 2/3 of $S/N$.
Measurements have shown that values of the noise and of the $S/N$ were stable and had only small variations with time over the period of the commissioning run at TIF.

\section{Tracking}\label{sec:tracking}
\begin{figure}
\begin{center}
\psfig{file=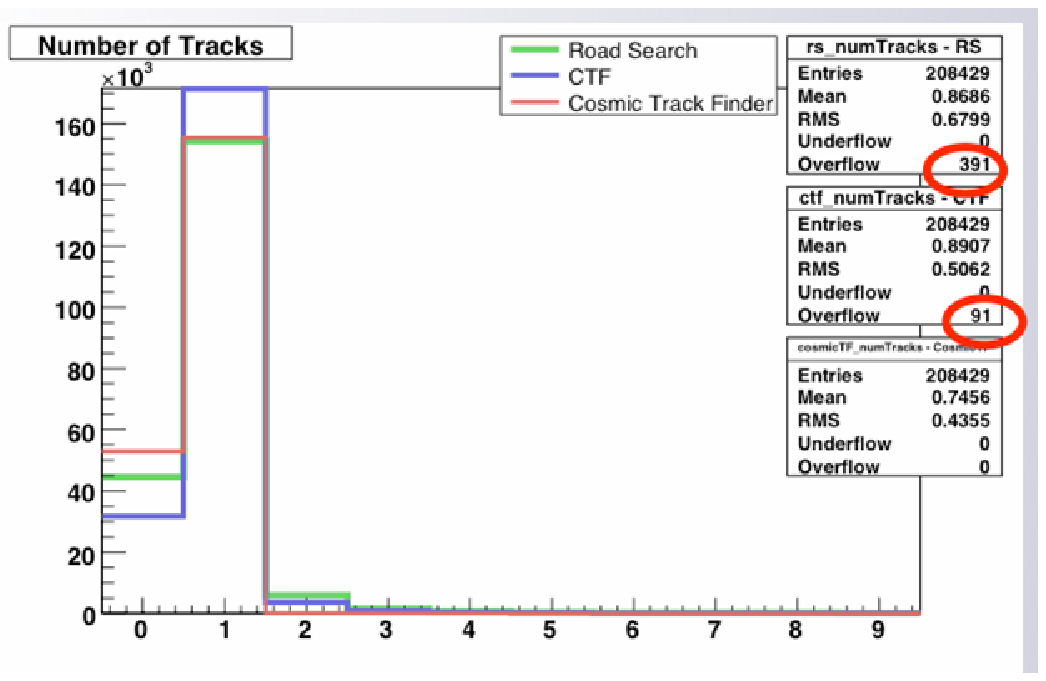,width=1.8in}
\psfig{file=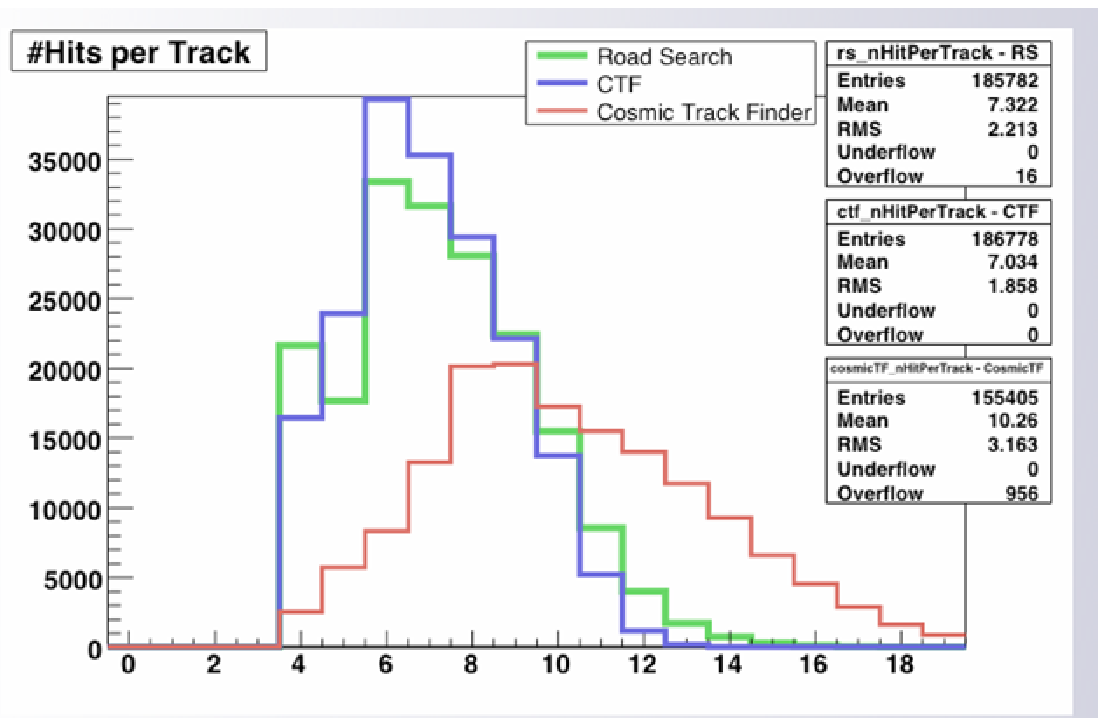,width=1.8in}\\
\psfig{file=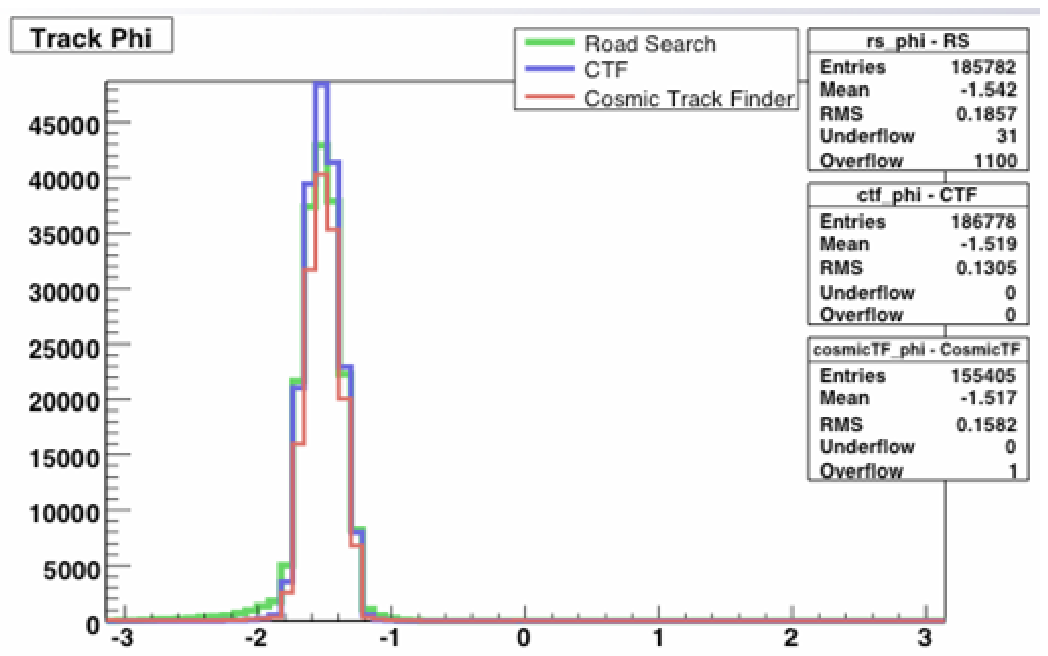,width=1.8in}
\psfig{file=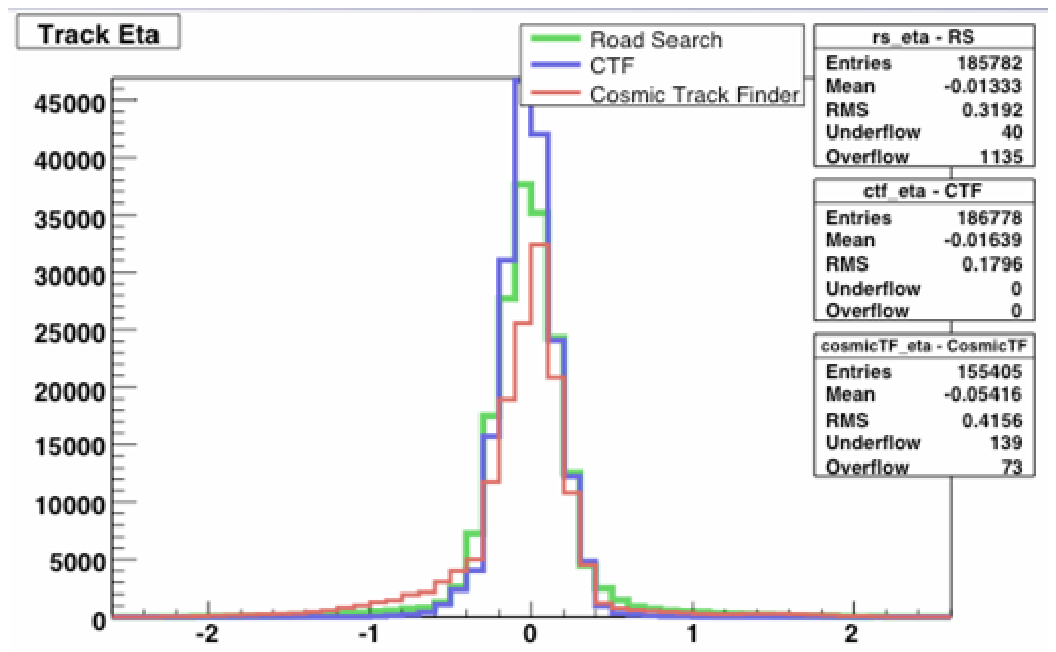,width=1.8in}
\end{center}
\caption{Track quantities for the three different track algorithms used at TIF and MTCC. From top left clockwise: number of tracks, number of reconstructed hits per track, track azimuthal angle, track pseudorapidity.}
\label{fig:tracking12}
\end{figure}

Three tracking algorithms were applied to TIF and MTCC data: the standard Combinatorial Track Finder~\cite{bib:ctf}, Road Search~\cite{bib:rs} and the Cosmic Track Finder~\cite{bib:cosmic}, specialized for single-track cosmic events.
 They use the reconstructed hits, i.e. position estimates based on clusters found in the modules of the tracker. In addition a reconstruction geometry describing the location of the modules and the distribution of passive material and condition information about the status of the different modules were needed. Modules known to be noisy were not taken into account.
In figure~\ref{fig:tracking12} some track quantities are shown for all algorithms. The Cosmic Track Finder has a larger number of hits per track because it treats hits from stereo detectors as two separate hits. A general agreement is seen between the different algorithms.

\section{Detector Efficiency Measurement}

\begin{figure}
\begin{center}
\psfig{file=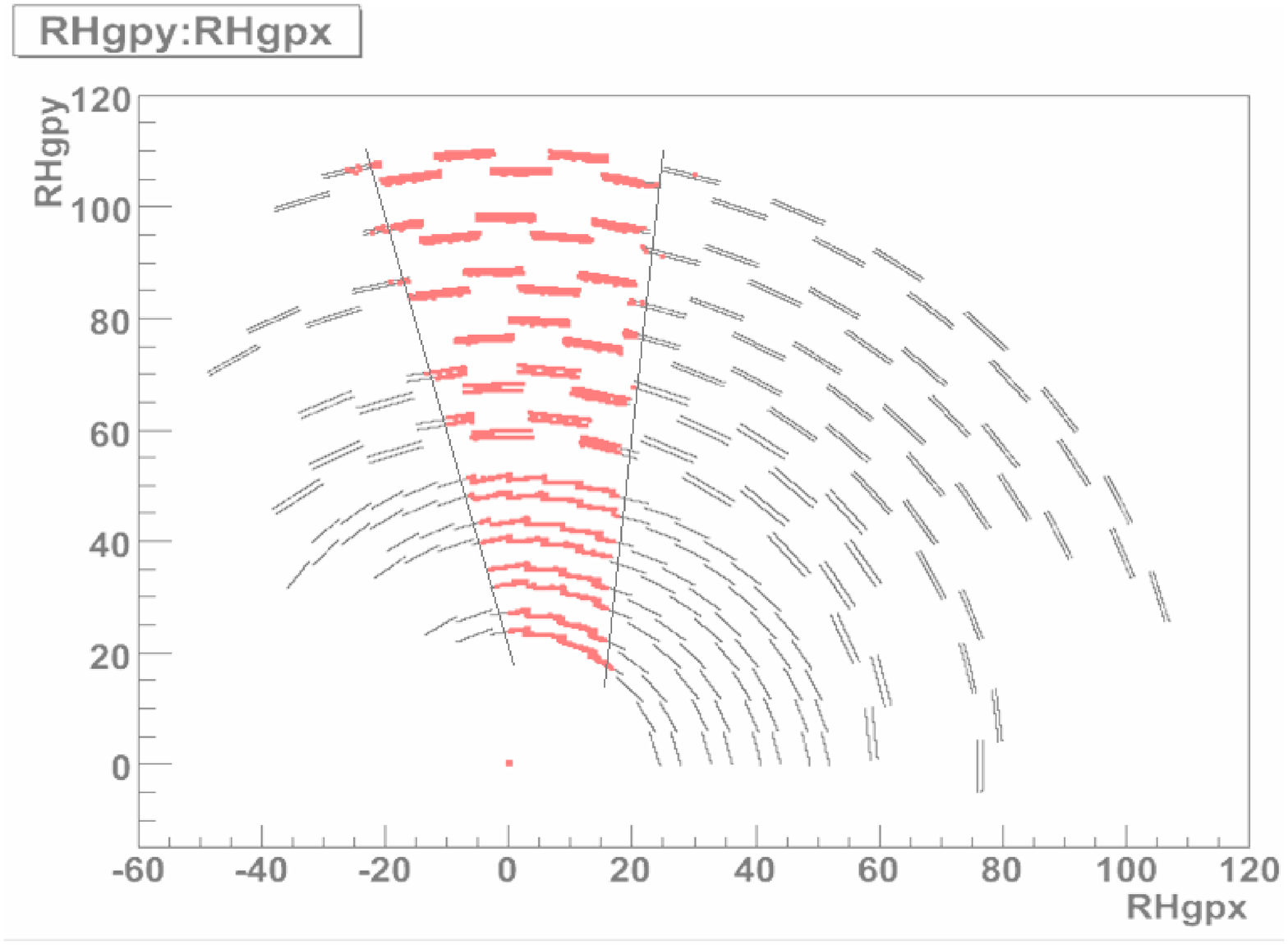,width=2.2in}
\psfig{file=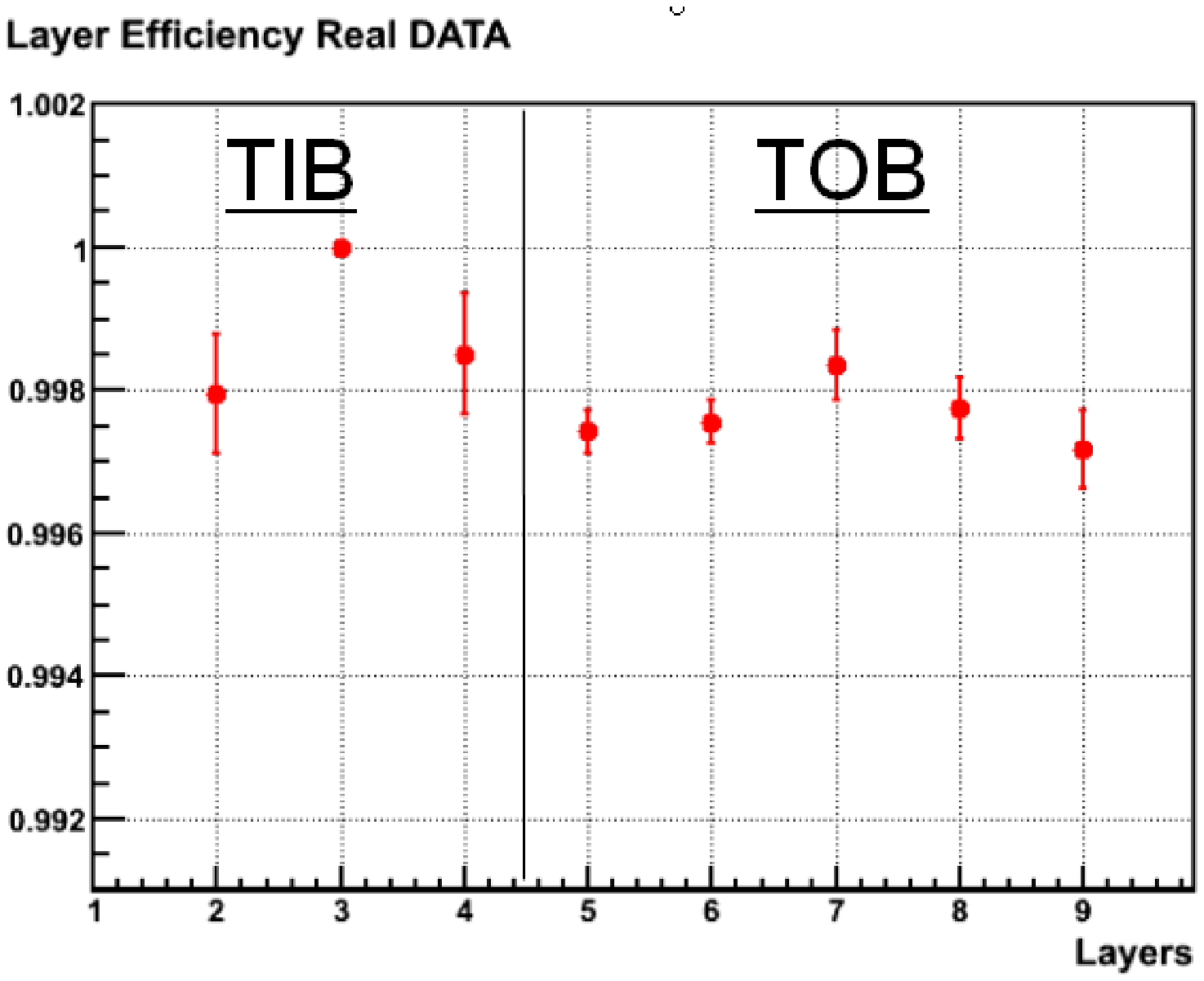,width=2.2in,height=1.6in}
\end{center}
\caption{Left: the shaded area between the two lines shows the selected region for determining the detector efficiency. Right: detector efficiency for TIB and TOB layers at the TIF.}
\label{fig:detefficiency}
\end{figure}

During the integration procedure the number of dead or noisy channels was determined to be low, around 0.2\% of the total~\cite{bib:tkintegration}. The tracker layer and module response efficiency was cross-checked using cosmic muon data taken at the TIF. For this the Combinatorial Track Finder was run and hits associated to the tracks were selected. For each layer (or module) crossed by the track the number of valid hits, $S$ and invalid hits, $B$ were computed, where valid means that the track built excluding that layer / module finds the hits in the expected position, within a certain range (dependent on the track / hit position uncertainty and the tracker alignment precision). The efficiency was then calculated as $S / (S+B)$. Events with only one track were selected in order to avoid high occupancy and tracks were selected almost perpendicular to the modules to avoid uncertainties in the module assignment during track propagation. The selected region for performing the study  and the measured efficiencies for TIB/TOB are shown in figure~\ref{fig:detefficiency}. The layer efficiency is larger than 99.7\% for both single sided and double sided layers.

\section{Gain Measurement}\label{sec:gain}

\begin{figure}
\begin{center}
\psfig{file=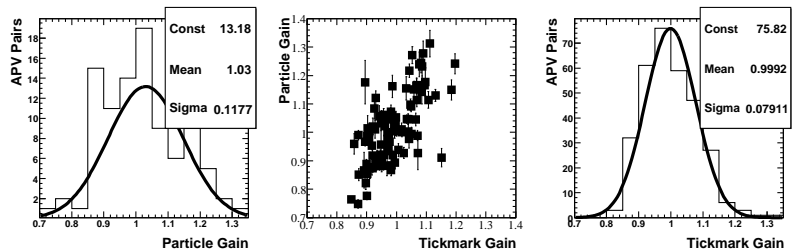,width=4.2in}
\end{center}
\caption{Distribution of gain inter-calibration constants for APV pairs calculated with the particle and tickheight methods as well as the correlation between the two.}
\label{fig:gaindistributions}
\end{figure}

Charged particles passing through the silicon material of the tracker release charge that translates into ADC counts assigned to the set of channels that make up a cluster. Non-uniformities both in the charge collection and in the readout chain affect the aplification and the linearity of the primary charge~\cite{bib:gainprediction}.
%
%A significant contribution to the gain non-uniformities comes from the Linear Laser Driver (LLD) on the AOH. The LLD has four gain settings, allowing a course gain equalization. The residual non-uniformity after the optimal settings are applied is still expected to be around 15\%.
%
Linearity and uniformity of the amplification (gain) across the channels of a silicon module is fundamental for the ultimate space resolution of these detectors. Also, the performance of particle identification technique with energy loss (Section~\ref{sec:energyloss}) depends both on the absolute calibration and on gain non-uniformities.
Two complementary methods are used to perform the inter-calibration of the APV pairs. The tickmark method uses a signal with constant height generated by each APV and consists in equalizing the height of it between modules.
The particle method uses the cluster charge of the hits (corrected for tracks' inclination) associated to reconstructed tracks and consists in equalizing the most probable value between different modules.  As such it takes into account also non-uniformities in the silicon, amplification chain preceding the Linear Laser Driver (LLD) and non-perfect synchronization of the readout.
The gain correction factors for both methods applied to MTCC data and their correlation on an APV pair basis are shown in figure~\ref{fig:gaindistributions}. A correlation between the methods is observed. The particle gain values are larger than the tick-height ones.

\section{Particle Identification with the Energy Loss Technique}\label{sec:energyloss}

\begin{figure}
\begin{center}
\psfig{file=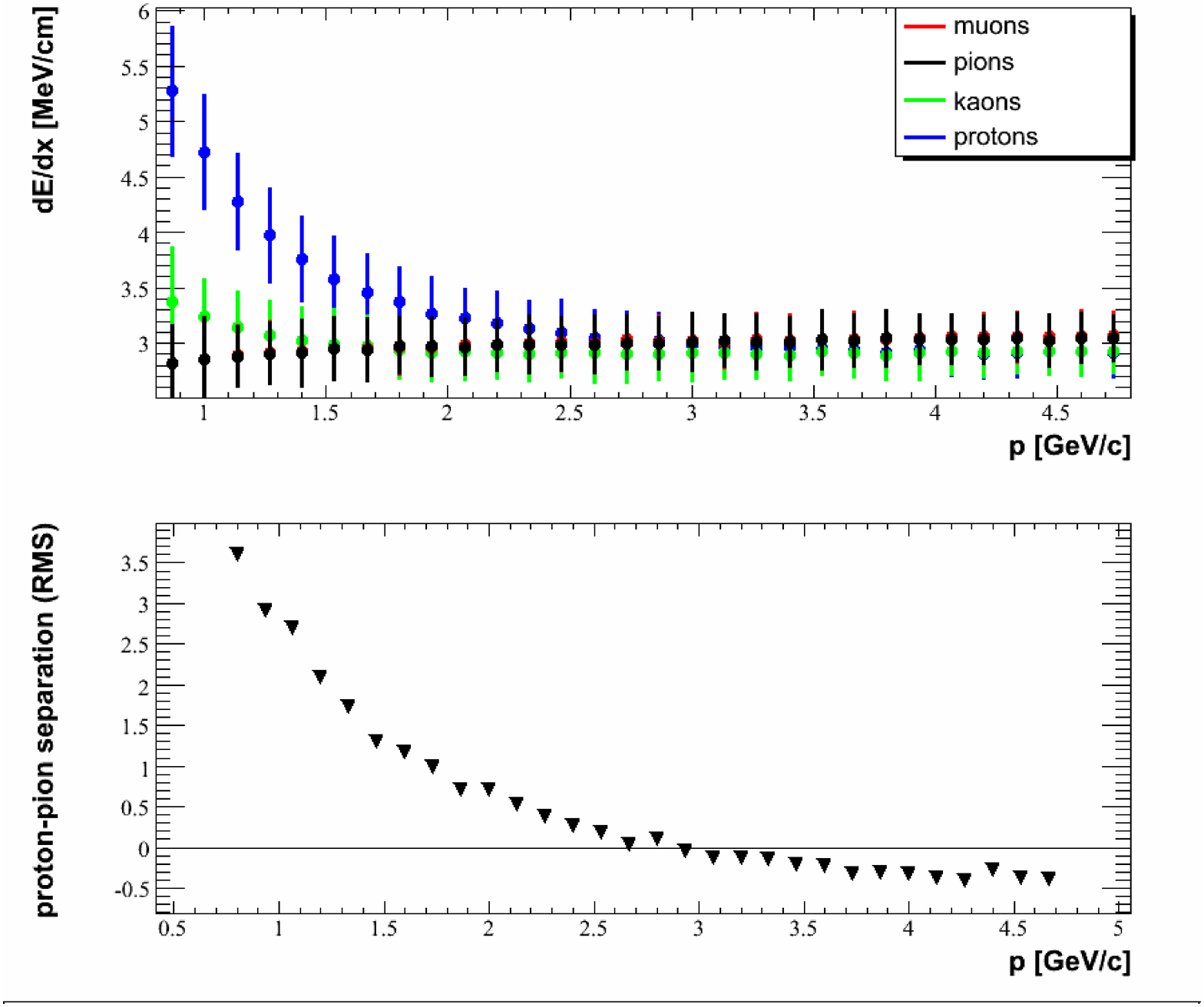,width=1.8in}
\psfig{file=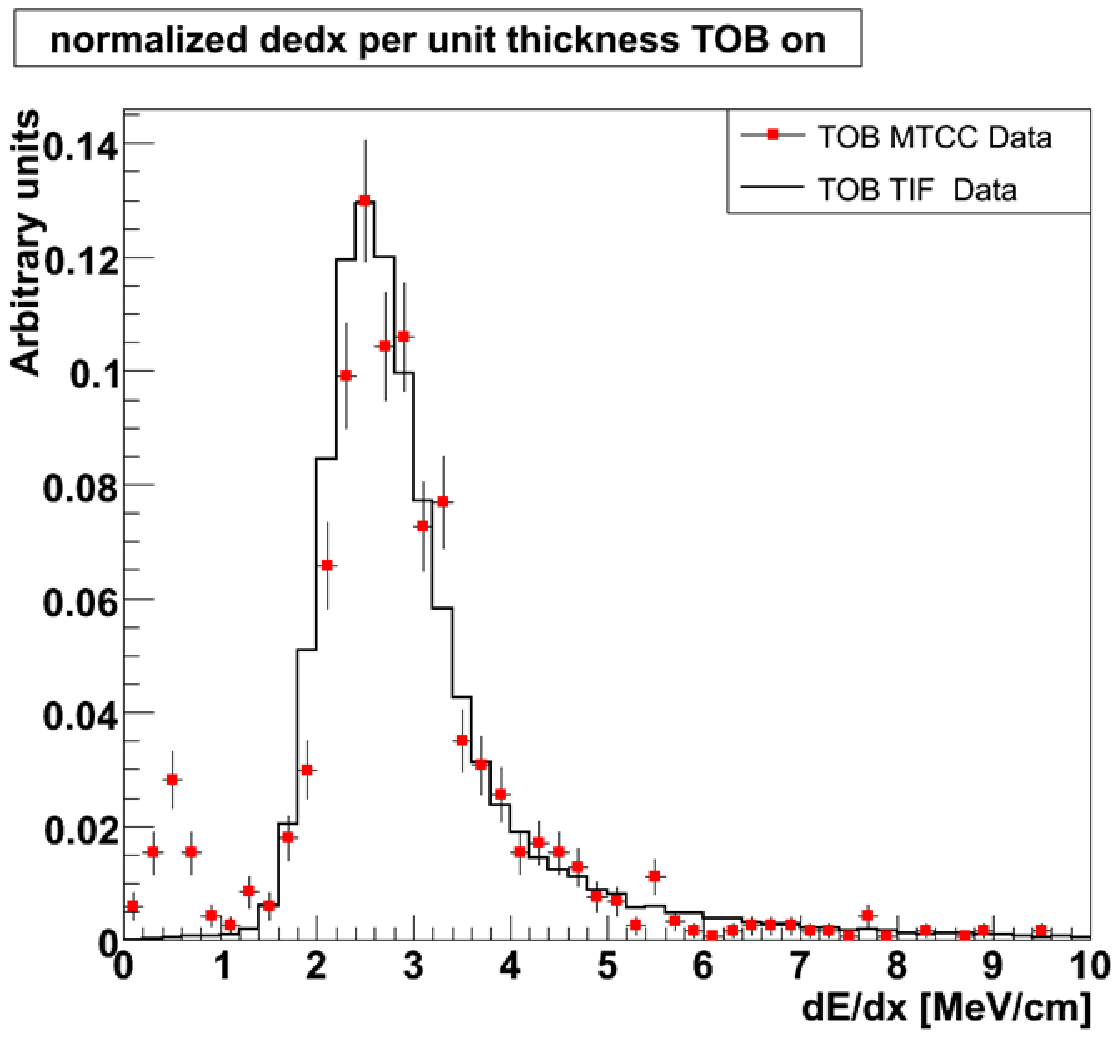,width=1.8in}
\end{center}
\caption{Left: most probable values for ${\rm d}E/{\rm d}x$ from a Landau fit and proton/pion separation as function of the track momentum. Right: Comparison of the ${\rm d}E/{\rm d}x$ for TIF and MTCC.}
\label{fig:dedx}
\end{figure}

The energy deposit into the silicon layers of the tracker can be used for particle identification. The signal height from a single microstrip (or pixel) is related to the number of electron-hole pairs created by the traversing particle in the bulk of the silicon module. The ${\rm d}E/{\rm d}x$ of the track can provide information for identification of electrons in jets and will be able to discriminate between different hadron species. This is important particularly for low $p_{\rm T}$ jets, for which correcting the energy of the proton using its mass instead of the pion mass makes a difference of around 1~GeV. Another important motivation for the development of ${\rm d}E/{\rm d}x$  measurements is that a large energy loss is one of the most characteristic signatures of long-lived massive charged particles\cite{bib:CMS_dedx}. The left side of figure~\ref{fig:dedx} illustrates the proton-pion separation at low momenta, where the separation is defined as the difference of the means over the square root of the sums of the squares of the two RMS for the $\log({\rm d}E/{\rm d}x)$ distributions of protons and pions.

% Simulated data from a particle gun and the full tracker simulation are used.
The right side of figure~\ref{fig:dedx} shows a comparison of ${\rm d}E/{\rm d}x$ measurements for the MTCC and the slice test of the TOB at the TIF. No difference is visible in this comparison. The ${\rm d}E/{\rm d}x$ in both cases is normalized to the path the particle travels in the silicon.

\section{Conclusions}

The commissioning run at the TIF and MTCC has been an important experience for the tracker. The tracking system has been successfully commissioned with local and global DAQ and operated together with all other subdetectors of CMS. The tracker perfomance has been excellent and a large sample of data has been gathered. Detailed offline studies are ongoing. Results of some of these studies have been presented.

%\section{Acknowledgments}
%This work is supported by the Belgian Interuniversity Attraction Pole P6/11.

\bibliographystyle{ws-procs9x6}
%\bibliography{ws-pro-sample}

\end{document}